\documentclass[preprint,12pt]{elsarticle}

\usepackage{amssymb}

\journal{Applied Mathematical Modelling}

\begin{document}

\begin{frontmatter}



\title{Efficiency in Micro-Behaviors and FL Bias}


\author[label1]{Kazutaka Kurihara}
\author[label2]{Yohei Tutiya}

\address[label1]{2-1-1, Tsuda-machi, Kodaira-shi, Tokyo, Japan}
\address[label2]{
1030, Shimo-Ogino, Atsugi, Kanagawa, Japan}

\begin{abstract}
In this paper, we propose a model which simulates odds distributions of pari-mutuel betting system under two hypotheses on the behavior of bettors: 1. The amount of bets increases very rapidly as the deadline for betting comes near. 2. Each bettor bets on a horse which gives the largest expectation value of the benefit. The results can be interpreted as such efficient behaviors do not serve to extinguish the FL bias but even produce stronger FL bias.
\end{abstract}


\end{frontmatter}


\section{History and Motivation}
The market efficiency hypothesis has been one of the most argued subjects in Economics (\cite{Bach,Fam} for standard references). It claims that if an arbitrage occurs in a market, news of this will spread so rapidly that on average people are unable to benefit. This notion has also been applied to the study of Pari-mutuel betting systems in the late 1900's. For example, if a horse wins too often in relation to its odds, it quickly becomes well known and people begin to bet more money on it. Hence, we may well expect its odds to decrease rapidly,  to the point where there is no more benefit in betting on this horse compared to any other. However, this is not always the case in real horse racing. It has been reported that expectation values of the benefit of betting on the favorites tend to be higher than the perceived benefit of betting on longshots (\cite{Gr} for the first report). This is called favorite-longshot (FL) bias. It should be noted here that the weak form of market efficiency is still observed under the bias. In other word, although the rate of return for favorites is large, it does not exceed one.  \\
\quad
 After the report \cite{Gr}, FL bias has been observed in many races all over the world. It is also notable that inverse FL bias, i.e. a phenomenon that rate of return  for longshots becomes higher than that for favorites, was observed in several parimutuel systems other than horse racing \cite{SobRyan}. In both FL and inverse FL bias, what is theoretically important is that the rates of return are different in favorites and longshots. This has been one of the central subjects in the study of the pari-mutuel betting system. Many preceding studies attempted to explain such bias by assuming the existence of non-efficient bettors. In \cite{Col, BruJoh}, for example, the bias is explained as an equilibrium of powers between two groups, one of which is constituted of efficient bettors while the other puts an importance on pursuing the leisure or thrill of betting. 
There also many studies which explain the bias in terms of misperception of the true underlying probabilities for winning \cite{ChadQua,ThaZie}. 
If the existence of non-efficient bettors or misperception of probabilities account for the bias, the next question will be how and why the ratio of such bettors or misperceptions are determined so coincidentally as to produce the real distribution of odds with weak efficiency being a possibility. 
 Before going to this stage, this study simulates what will happen if all the betters behave efficiently. The result suggests that the FL bias occurs even if all the bettors behave to maximize their expectation values of their bets. The bettors in the model bet their money on horses whose odds are large in relation to their probabilities of winning. And the odds fluctuate over time with such bets until the race begins. Conditions for the simulations will vary in several ways. We will  change how efficiently bettors behave. It will also be changed how strong favorites are in a race. We will then change how rapidly bets accumulate over time. \\
 \quad
It should be noted here that there are already some studies which explain FL bias 
by considering how odds tables are driven by wagering over time.\ 
Some of the most successful examples will be Ali \cite{Ali}, Blough \cite{Blo}, Brown and Lin \cite{BroLin}.\ 
In their studies, bettors hold heterogeneous subjective probabilities about the true winning probability of each horse. 
Contrary to that, in our model, the beliefs of bettors are homogeneous. But the attitudes for efficiency fluctuate randomly.\ 
According to our results, FL bias is even amplified when bettors behave too efficiently. 

\section{Modeling}
 Let us consider the win betting system in a race consists of \(N\) horses. We denote the ratio of the amount of bets for the \(i\)-th horse to that of all horses at time \(t\) as \(V_i(t)\). In pari-mutuel betting system, the odds for this horse is \((1-\alpha)/V_i(t)\), where the constant \(\alpha\) is the rate of track take fixed in each racetrack. Since \(V_i(t)\) varies over time, the odds table also varies over time. In most racetracks, odds tables changes at one or two minute intervals. And, the change of the odds table becomes one of major reasons to promotes changes in \(V_i(t)\) because people make their decisions based on the odds table. According to this mechanism, \(V_i(t)\) changes over time until the race starts.
These are, at this time, the general setting of the pari-mutuel system. 
 Now we introduce a positive constant \(p_i\), which represents a rational value of \(V_i(t)\). And we regard \(p_i/V_i(t)\) to represents how much it is efficient to bet on \(i\)-th horse. That is to say, the larger \(p_i/V_i\) , the more the \(i\)-th horse appealing to efficient bettors. 
If the track take is zero, \(p_i\) can be regarded as the true probability that \(i\)-th horse wins, then, \(p_i/V_i\) is nothing but the rate of return of betting on \(i\)-th horse. One might wonder that, if it were roulette, bettors are in advance informed of the probability that each number will come up. But, in the case of horse racing, people do not know any such exact values to judge the efficiency beforehand. More precise explanations on this issue will be given in section 2.3. 
 Another remarkable tendency of efficient bettors is that they like to put off their bets for as long as possible. To reflect this pattern, our model only deals with the time evolution of bets in the last  \(20\) minutes before races start. Thus, we arrive at the following model.
 \begin{itemize}
\item[1.]
Let  \(t\ (t=-20,-19,\cdots,0[{\rm min}])\) denotes time after a race starts.  
The total amount of minimum unit of bets in each race is fixed to be 200,000.\
And the rate of bets finished before time \(t\) will be denoted as \(F(t)\). Hence, \(F(0)=1\) is assumed. We will also use \(f(t)=F(t)-F(t-1)\) to represent the rate of bets which are accumulated between \(t-1\) and \(t\). 
\item[2.]
The number of horses in each race is fixed to be \(N\) . And each horse equips a constant which will be referred to as consensus probability \(p_i\ (i=1,2,\cdots,N)\) \cite{Blo}. We assume \(p_1+ p_2+ \cdots+p_N=1\) and \(p_1\geq p_2\geq \cdots\geq p_N\) for convenience.  
\item[3.]
At time \(t\), the all \(N\) horses are sorted according to the magnitude of \(p_i/V_i(t)\). Then we introduce a probability \(q_n\) with which a bet is made on a horse having the \(n\)-th largest magnitude. \(\{q_n\}_{n=1,2,\cdots,N}\) will be denoted as the preference probability distribution, where \(q_1+q_2+ \cdots+q_N=1\) and \(q_1\geq q_2\geq \cdots\geq q_N\) are assumed. 
\end{itemize}
Now we proceed to choose the parameters of this model. 
\subsection{The shape of \(F(t)\)}
\begin{figure}
\begin{center}
\includegraphics[width=.9\linewidth]{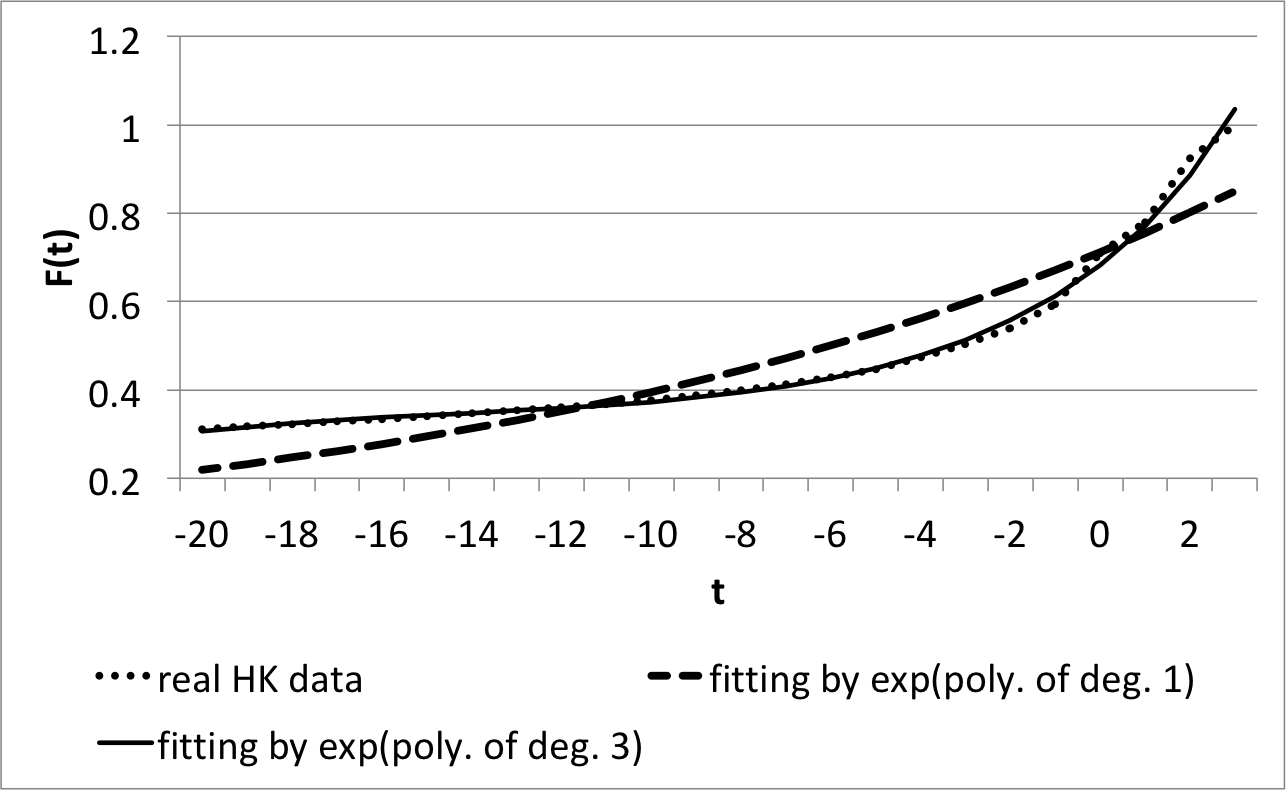}
\caption{\label{fig:invest_final} an average of \(F(t)\) in HK racetrack and exponential fittings.}
\end{center}
\end{figure}

If the bettors are genuinely efficient, they will not make any bet until the last minute before each race starts. For example, \cite{GraMac} reports that 40\% of the total bets rash to the last few minutes before a race starts.\ According to this idea, the most suitable distribution is a delta function, i.e. \(f(t)\) takes positive value only when \(t=1\) and \(f(t)=0\) otherwise. But, in reality, ``the last minute" can be different from bettor to bettor. So there must be a fluctuation from the delta function. 
Here we exhibit an example of \(F(t)\) in a real race track in FIG.~\ref{fig:invest_final}.\ 
The real line represents the average \(F(t)\) in the HongKong racetrack \cite{Hong} in the period from October 2012 to May 2014, where the average is taken over entire 2000 races during the period. The horizontal axis represents the time \(t\). It should be noted here that numbers of bettors in real data  change until \(t=2\) or 3 probably because of a lag of publication.  
 The regression curve of dotted line is of the form \(\exp(c_0+c_1t)\) and the dashed one \(\exp(c_0+c_1t+c_2t^2+c_3t^3)\).\
It would be an interesting future study to investigate the mechanism that generates the shape of \(F(t)\). Instead of sticking to the exact shape, in this paper, we will change \(f(t)\) in several ways. As for the speed of growth, we will try exponential growth, constant growth and exponential decay. Another important factor is the initial value \(F(-20)\) . We will consider \(F(-20)=0.35\) and 0.01. Up to Section 5, we only deal with exponentially growing \(f(t)\) with \(F(-20)=0.35\), which gives the nearest shape for HK data in Fig.\ref{fig:invest_final}. Henceforth in this paper the HongKong data will be regarded as standard for setting parameters.\\
 \quad 
  It should be noted that very few pari-mutuel racetracks disclose data on \(F(t)\). And the HongKong racetrack is one of the most notable examples of world famous racetracks which openly disclose their data on \(F(t)\).
\subsection{Choice of \(N\)}
We set as \(N=14\). This is not only because the number of horses in a HongKong race is 14, but also because most races in the world are held with 10 to 16 horses.
\subsection{Choice of consensus probability
}
As stated in the rule, we associated the consensus probability \(p_i\) with each horse. This might seem strange because people do not know the probability that each horse will win. However, each bettor must possesses his/her own valuation on each horse and the average of these valuations must exist as \(p_i\). Since it is just an average, it can differ from valuations of individual betters. And the belief in \(p_i\) can fluctuate from time to time. The value of \(p_i\) itself might also fluctuate over time. In this paper, we take a viewpoint that such variance and fluctuation in \(p_i\) are rounded to choice of preference distribution \(\{q_n\}\) . Thus, we set,
\begin{eqnarray}
\displaystyle p_i=C_a\exp\left(-ai\right)\label{ref_prob_eq}
\end{eqnarray}
where \(C_a\) is the normalization constant to make \(\sum_{j=1}^{14}p_j=1\) hold. We carried out the simulations with \(a=0.05,\, 0.1,\, 0.3,\, 0.5,\, 1.3\). Judging from the frequency of winning of favorites in the data, \(a\) is estimated to be \(0.3\sim0.5\) in real HongKong races. 
\begin{figure}
\begin{center}
\includegraphics[width=.7\linewidth]{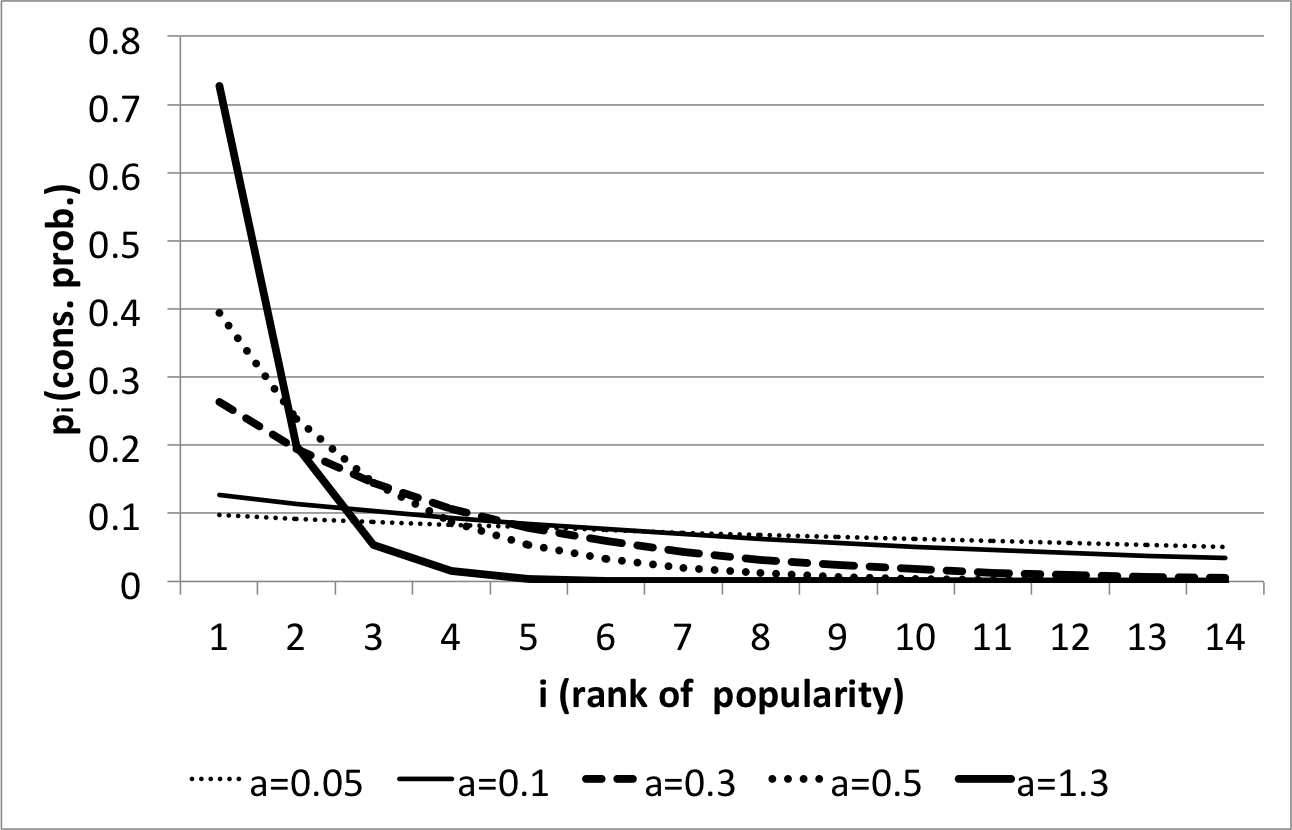}
\caption{\label{fig:ref_prob} 
Shapes of \(\{p_i\}\) for various values of \(a\).}
\end{center}
\end{figure}
\subsection{Choice of preference probability}
\begin{figure}
\begin{center}
\includegraphics[width=.7\linewidth]{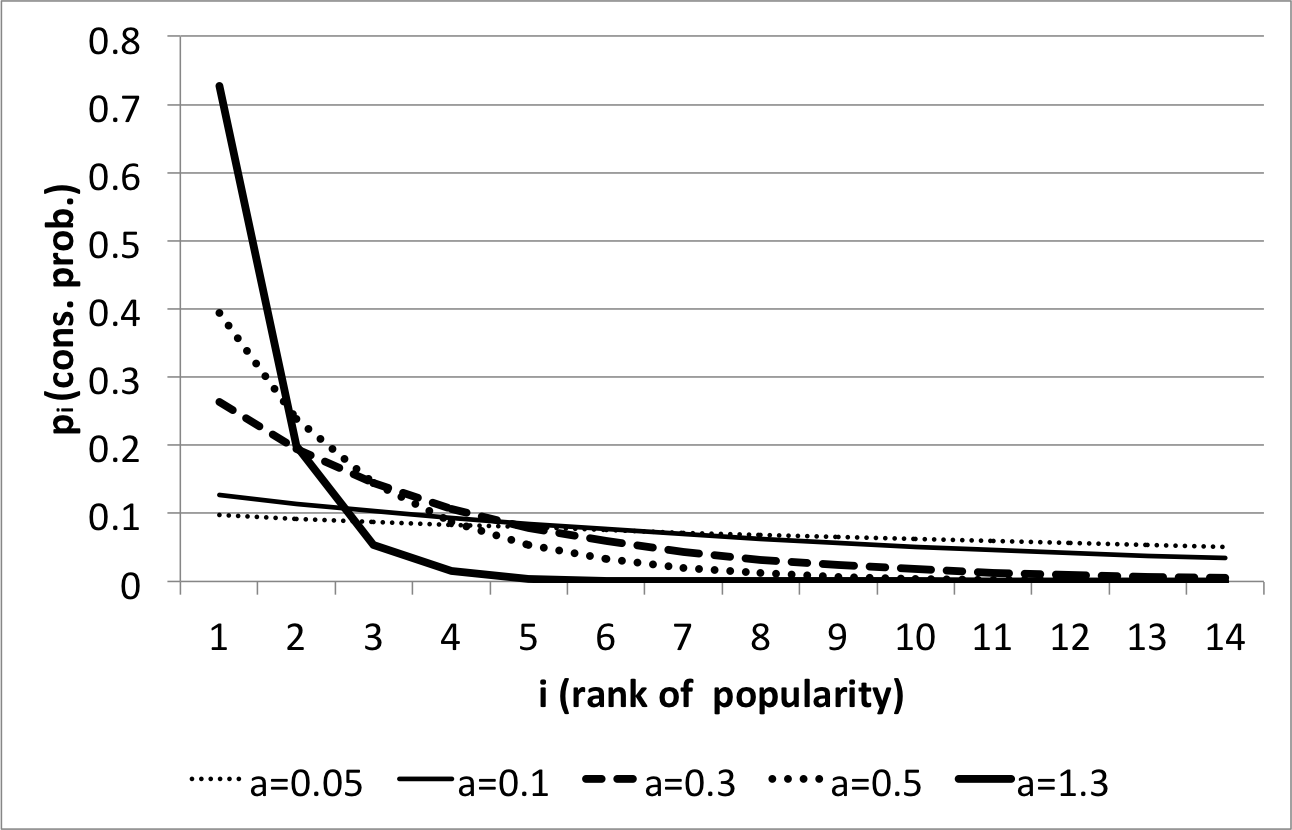}
\caption{\label{fig:pref_prob} 
Shapes of \(\{q_n\}\) for various values of \(b\).}
\end{center}
\end{figure}

As for \(q_n\), the only assumption one can naturally make is that \(q_n\) decreases with \(n\). To make variety of \({q_n}\) wide enough, we use the following function.
\begin{eqnarray}
q_n=D_b\left(\frac{15-n}{14}\right)^b
\end{eqnarray}
where \(D_b\) is a normalization constant which make \(\sum_{j=1}^{14}q_j=1\) hold.
We set \(b=1/400, 1/54, 1/20, 1/7, 1/4, 1/2, 1, 2, 4, 7, 20, 54, 400\). The shape of \(\{q_n\}\) for these values of \(b\) are listed in FIG. \ref{fig:pref_prob}.
\section{Results of simulations}
We carried out 2000 simulations for each set of parameters. 
The source cords and HongKong data of \(F(t)\) are kept at \\
https://github.com/qurihara/EconomicSimulation.\ 
\subsection{Preference distribution and FL bias}
\begin{figure}
\begin{center}
\includegraphics[width=.8\linewidth]{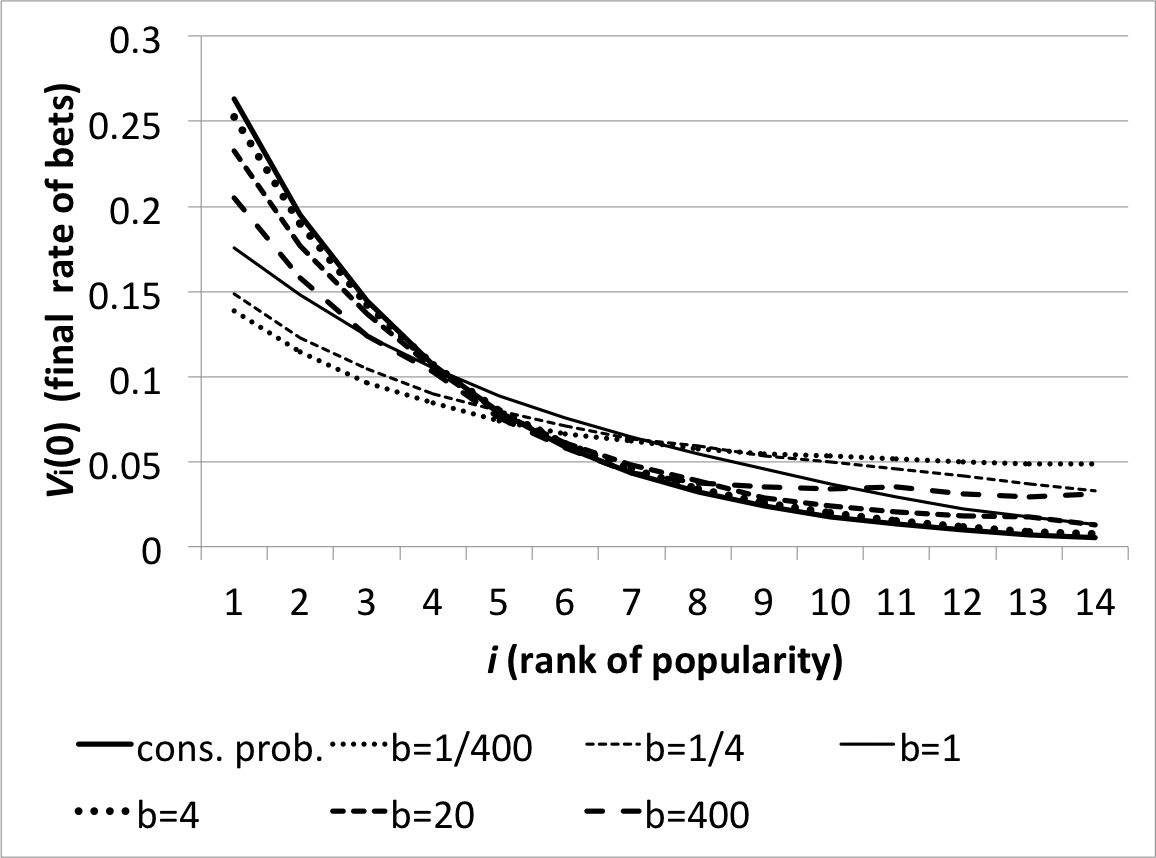}
\caption{\label{fig:standard} 
popularity rank and rate of bets compared to consensus probabilities}
\end{center}
\end{figure}

We first present a result of the case \(a=0.3\) in FIG.\ref{fig:standard}. The accumulation of bets is exponentially growing type and initial rate of bets \(F(-20)\) is set to be 0.35.
In FIG.\ref{fig:standard}, \(b=400\) implies that all the bettors bet their money on the most beneficial horse. While, the almost uniform distribution, \(b=1/400\), implies all the bettors bet their money on a randomly chosen horse regardless whether the horse is beneficial or not. It is not surprising that a FL bias appeared in the latter case, because people bet their money on longshots in equal probability as on favorites. It is truly astonishing that a strong FL bias is formed in \(b=400\) case. The bias seems weakest around \(b=4\).\ 
\begin{figure}
\begin{center}
\includegraphics[width=.9\linewidth]{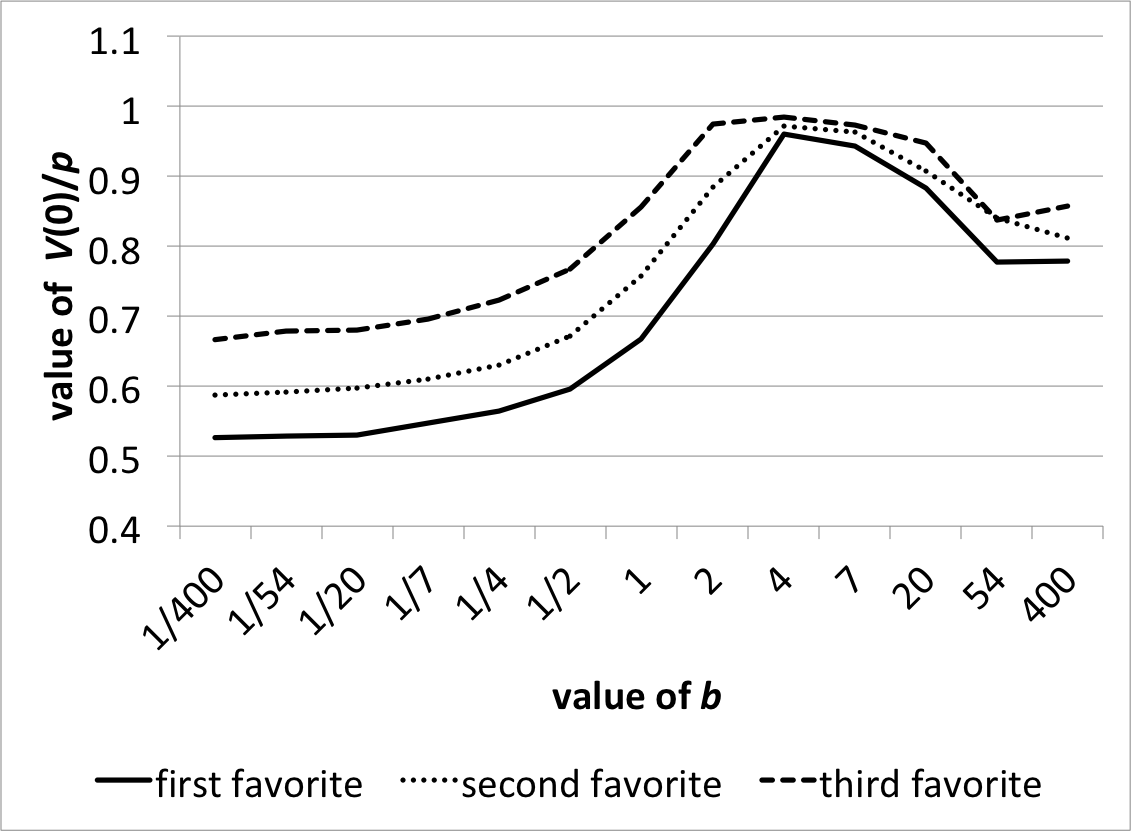}
\caption{\label{fig:1to3} 
\(V_i(0)/p_i\) for top 3 favorites.}
\end{center}
\end{figure}
Seeing FIG.\ref{fig:standard}, it is sufficient to compare \(V_1\) with \(p_1\) to measure the bias. Hence, in FIG.\ref{fig:1to3},  \(V_i(0)/p_i\) for \(i=1,2,3\) are exhibited, i.e., for top 3 favorites. In this figure, \(V/p<1\) means FL bias. 
The result suggests that there exists a value of \(b\) which maximize \(V_i/p_i\). The simulation claims that FL bias occurs even if all the betters behave efficiently. Moreover, it claims that a bias can even be weakened when there is proper amount of inefficient betters. 
\subsection{Uniformness of consensus distribution and FL bias}
\begin{figure}
\includegraphics[width=.9\linewidth]{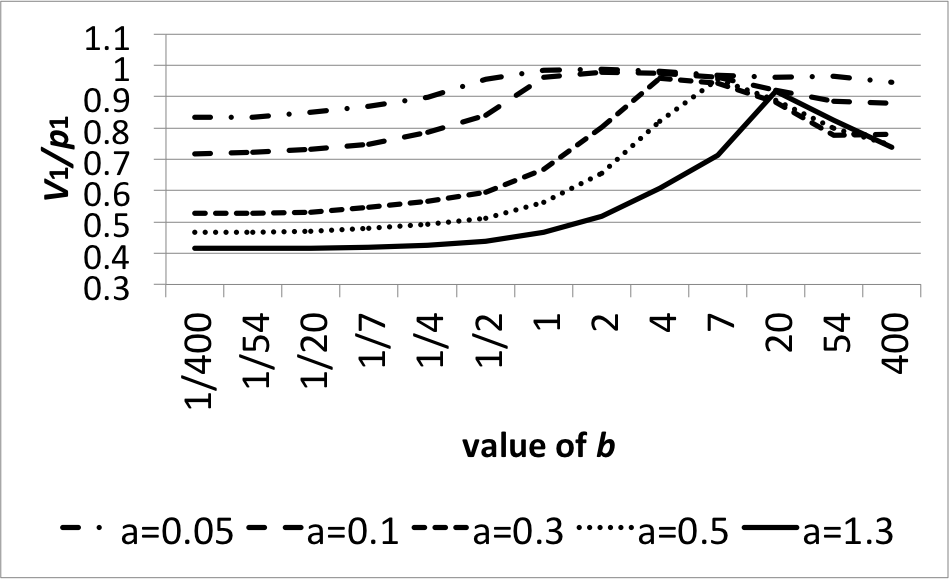}
\caption{\label{fig:p_change} How FL bias depends on consensus distributions.}
\end{figure}

Now, it is natural to change consensus distribution because bias can not happen when neither favorite nor longshot exists, i.e., consensus probability distribution is uniform. The bias must depend on how much stronger favorites are than longshots. Seeing the equation (\ref{ref_prob_eq}), the distribution tends to uniform distribution when \(a\) goes to \(0\). While, as \(a\) becomes greater than \(0\), the favorite becomes stronger. FIG.\ref{fig:p_change} represents results of
simulations for \(a=0.05,\ 0.1,\ 0.3,\  0.5,\ 1.3\). 
As the result shows, basically, the more outstaning the favorite, the stronger the bias becomes. Another notable feature is how the bias depends on preference distribution. Every time we fix \(a\), there exists a value of \(b\) which best weaken the FL bias. And such \(b\) becomes greater as \(a\) becomes greater. That is to say, the stronger the favorite becomes, the more we should bet efficiently to eliminate the FL bias. But bias does not disappear regardless. And yet another mystery is that the bias is again enforced when \(b\) exceed the best value.  
\subsection{Rapidity of accumulation of betting and FL bias}
\begin{figure}
\begin{center}\includegraphics[width=.9\linewidth]{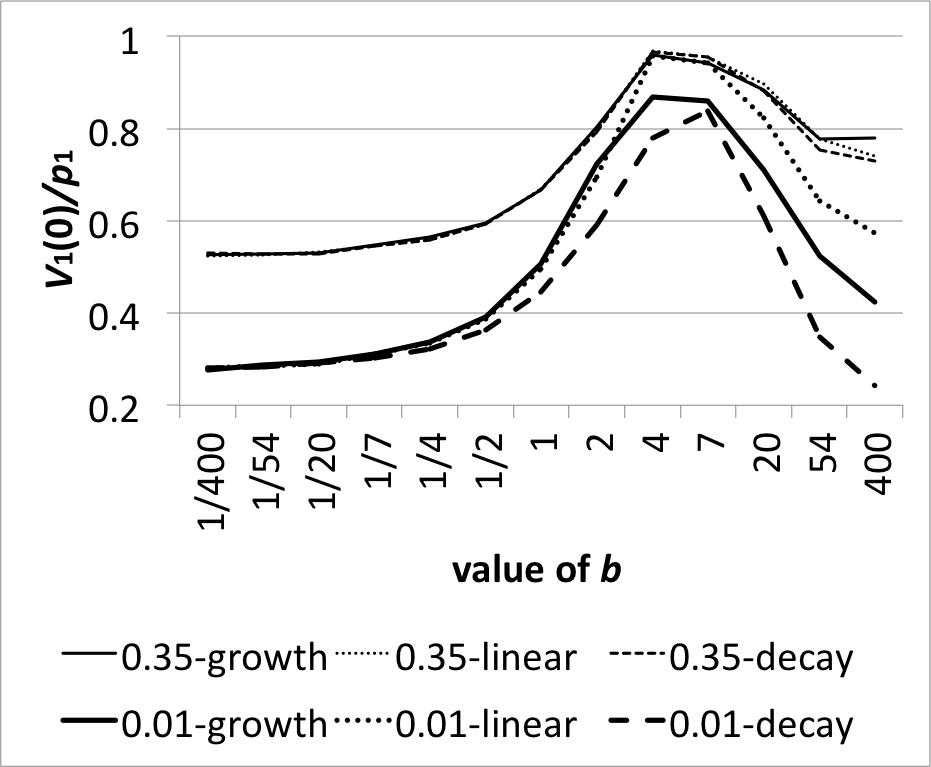}
\caption{\label{fig:invest_change} How \(V/p\) changes with \(f(t)\)}
\end{center}
\end{figure}

Since FL bias is a kind of over/under-estimation, it must depend on the amount of bettors who are making estimations at time \(t\). Hence, in this section, we will change the time dependence of the number of bettors. We introduced two viewpoints in section 2.1. First, we change the percentage of what we call pre-bettors, i.e., the bettors already finished their bets more than 20 minutes before a race start. In real Hong-Kong races, a typical percentage of pre-bettors is 35\%. But it is apparent that a percentage of pre-bettors has a large effect on the final distribution of bettors. Hence we not only examined 35\%, but also a condition that the percentage of pre-bettors is very small, i.e., 1\%. In both cases, we made the odds distributions of pre-bettors correspond to consensus probability distribution with a certain amount of random fluctuation. Second of all, we changed the rapidity of the growth of the numbers of bettors during the last 20 minutes. We examined exponential growth, linear growth and exponential decay. In all, we examined \(2\times3=6\) models for the growth of numbers of bettors. The mathematical expression for each model becomes unique due to the conditions \(F(0)=1\) and F(-20)=0.35\ (or 0.01). \\
 \begin{table}
	\centering
	\begin{tabular}{|l|l|l|l|}
		\hline
abbreviation&growth of \(f(t)\)&pre-bettors&expression of \(F(t)\)\\
\hline
0.35-growth&exponential growth&35\%&\(0.35^{-t/20}\)\\
\hline
0.35-linear&linear growth&35\%&0.0325t+1\\
\hline
0.35-decay&exponential decay&35\%&\(1.35-0.35^{1+t/20}\)\\
\hline
0.01-growth&exponential growth&1\%&\(0.01^{-t/20}\)\\
\hline
0.01-linear&linear growth&1\%&0.0495t+1\\
\hline
0.01-decay&exponential decay&1\%&\(1.01-0.01^{1+t/20}\)\\
\hline
	\end{tabular}
	\caption{models for  \(f(t)\).}
	\label{expfit}
	\end{table}
FIG.\ref{fig:invest_change}  represents how FL bias depends on the rapidity of growth of bettors. The vertical axis again represents \(V_1/P_1\). The parameter \(a\) is set to be \(0.3\).
As is shown, FL bias appears more radically when the percentage of pre-bettors is 0.01. Now we have to be mindful that the pre-bettors bet in accordance with the consensus probability  distribution. This might partially explain why \(V_1/p_1\) becomes higher for 1\% case than for 35\% case. 
Another outstanding feature is that the bias is strongest in 0.01-deay model. 
We do not yet know any quantitative explanation for these result.  
\section{Discussion and future works}
The model in this paper assumes that each bettor behaves to make \(V_i\) near to \(p_i\). However, \(V_i\) becomes greater than \(p_i\) for longshots and smaller for favorites. This can be interpreted as a FL bias. 
Since it is a purely statistical model with homogeneous bettors, the reason for this bias should exist in macro-micro interaction, i.e., sorting of horses. We believe it is interesting that such simple and uniform rules in micro behavior result in unexpected macro statistics. 
It is also observed that the strength of bias depends on consensus probability distributions and how bets accumulate over time. The more favorites outstand, the more bias get stronger. The bias becomes stronger if \(f(t)\) has points where it increases rapidly.  
 Qualitatively speaking, a minimum unit of bet for longshots makes a larger disproportionate effect on odds distribution than for favorites because the number of bettors in longshots is smaller than those in favorites. For example, assume that the number of bettors for longshots is 1 and for favorites 10. Then, \(V\) for longshots is \(1/11\) and favorites 10/11. Now assume that one bet is added to longshots. Then, \(V\) for longshots becomes \(2/12\). If it is added to favorites, \(V\) becomes \(11/12\). In this example, it is apparent that the change in \(V\) is much greater in longshots. But we yet do not know whether this explains our results enough. Due to the sorting process, it is difficult to give mathematical analysis. Probably, analysing \(N=2\) case will be a first future work. A master equation for \(N=2\) case is exhibited in Appendix.\\
\quad
Another branch for this study is that it can serve to analyze real systems similar to pari-mutuel system. For example, it simulates games such as roulette and baccarat where consensus distributions are disclosed to bettors. Another type of application is for growth of queues at counters in shops or gates on highways. In these settings bettors are interpreted as customers. each customer choose a queue to join, judging from the consensus length of the queue. Favorites now implies counters or gates which have shorter consensus length. In the field of stochastic process, our model is a kind of anomalous walk where the drifts can vary depending on the position of walkers.\\ 
\quad In terms of pari-mutuel betting systems, the most necessary modification for the model will be to introduce a track take. In real horse races, efficient bettors will bet their money only when the benefit is positive even if the track take is considered. That probably means we should set a threshold for betting in terms of \(V_i/p_i\). 

\appendix
\section{Master equation for \(N=2\) case}
Let \(P(\cdot ,t)\) denote the probability density function for \(V_1(t)\). Then, the state \(k<V_1(t)<k+dk\) occurs with the probability \(P(k,t)dk\).\
The aim of this appendix is to show that the below is the time evolution equation of \(P(k,t)\) for \(N=2\) case.\ 
\begin{eqnarray}
&&P(k,t)=\int_{-\infty}^{p}P(s,t-1)\, \phi\left(k-s;\, \frac{(s-q)f(t)}{F(t)},\, \frac{q\overline{q}f(t)}{mF(t)^2}\right)ds\nonumber\\
&&\hspace{2em}+
\int_{p}^{\infty}P(s,t-1)\, \phi\left(k-s;\, \frac{(s-\overline{q})f(t)}{F(t)},\, \frac{q\overline{q}f(t)}{mF(t)^2}\right)ds.
\label{master}\\
&&\displaystyle\phi(x;\, \mu,\sigma^2)=
\frac{1}{\sqrt{2\pi}\sigma}\exp\left(-\frac{(x-\mu)^2}{2\sigma^2}\right)
\nonumber
\end{eqnarray} 
Here, we defined \(q=q_1,\  \overline{q}=q_2(=1-q)\) and \(p=p_1\) for simplicity. 
\(m\) is the total number of bets in a race fixed as 200,000 in this paper.\ 
\(\phi(\cdot\,;\mu,\sigma^2)\) is the probability density function for \({\rm N}(\mu,\sigma^2)\), the normal distribution with the average \(\mu\) and variance \(\sigma^2\).\\ 
\quad First notice for deriving (\ref{master}) is that, if \(V_1(t-1)<p\), the amount of bets accumulating to the favorite at time \(t\) yields \({\rm B}(mf(t), q)\). Here, \({\rm B}(n, r)\) denotes the binomial distribution with \(n\) being total number of experiments and \(r\) the probability of a successful result. Then, since \(mf(t)\) is large, \({\rm B}(mf(t), q)\) is approximated by \({\rm N}(mqf(t),\ mq\overline{q}f(t))\).\ 
Then, we interpret this observation in terms of \(V_1(t)\). Suppose \(V_1(t-1)\) increases from \(y\) to \(y+z\) at time \(t\), i.e. \(V_1(t)=y+z\).\ 
This implies that \(m(zF(t)-yf(t))\) bets are made at time \(t\).\ 
Hence, the increment \(z\) yields \({\rm N}((y-q)f(t)/F(t),\ f(t)q\overline{q}/mF(t)^2)\).\ 
We can analyse the \(V_1(t-1)\geq p\) case in the same manner. In this case, \(z\) yields \({\rm N}((y-\overline{q})f(t)/F(t),\ f(t)q\overline{q}/mF(t)^2)\).\\ 
\quad Now, we proceed to consider a recurrence equation for \(P(k,t)\).\ One way to reach the state \(V_1(t)=k\) is that \(V_1\) increases by \(k-s\) after \(V_1(t-1)=s\) occurred.\ 
The probability density of this is a multiple 
\[P(s,t-1)\cdot \phi\left(k-s;\, \frac{(s-q)f(t)}{F(t)},\, \frac{q\overline{q}f(t)}{mF(t)^2}\right), 
\] 
if \(s\) is less than \(p\). 
The r.h.s. of the equation (\ref{master}) is nothing but an integration of such multiples for various values of \(s\).\ 
\section{References}



\end{document}